\documentclass[aps, prd,10pt,notitlepage,nofootinbib,superscriptaddress,showkeys,twocolumn]{revtex4}
\usepackage{amstext,amsmath,amssymb,amsfonts,bbm}
\usepackage[latin1]{inputenc}
\usepackage{graphicx}
\usepackage{hyperref}
\usepackage{epstopdf}
\usepackage{amsthm}
\usepackage{pstricks}
\usepackage{tocvsec2}

\usepackage{color}

\usepackage{hyperref}

\def\beq{\begin{equation}}
\def\be{\begin{equation}}
\def\ee{\end{equation}}
\def\bes{\begin{eqnarray}}
\def\ees{\end{eqnarray}}

\theoremstyle{definition}
\theoremstyle{definition}
\theoremstyle{definition}
\theoremstyle{definition}
\theoremstyle{definition}
\theoremstyle{definition}

\begin{document}
\maxtocdepth{subsection}
%%%%%%%%%%%%%%%%%%%%%%%%%%%%%%%%%%%%%%%%%%%%%%%%%%%

\title{\large \bf Comment on `Lost in Translation: Topological Singularities in Group Field Theory'}

\author{{\bf Matteo Smerlak}}\email{smerlak@cpt.univ-mrs.fr}
\affiliation{Centre de Physique Th\'eorique, Campus de Luminy, Case 907, 13288 Marseille Cedex 09 France}

\date{\small\today}

%%%%%%%%%%%%%%%%%%%%%%%%%%%%%%%%%%%%%
\begin{abstract}\noindent
Gurau argued in \href{http://arXiv.org/abs/arXiv:1006.0714}{[arXiv:1006.0714]} that the gluing spaces arising as Feynman diagrams of three-dimensional group field theory are not all pseudo-manifolds. I dispute this conclusion: albeit not properly triangulated, these spaces are genuine pseudo-manifolds, viz. their singular locus is of codimension at least two. 

\end{abstract} 
%%%%%%%%%%%%%%%%%%%%%%%%%%%%%%%%%%%%%%

%MSC numbers: 81T45 (principal), 57M20, 81T25, 83C45 (secondary)
\keywords{group field theory, pseudo-manifold.}
\maketitle
%\tableofcontents

%%%%%%%%%%%%%%%%%%%%%%%%%%%%%%%%%%%%%%

Group field theory (GFT) is a tentative framework for quantum gravity, introduced by Boulatov in \cite{Boulatov:1992vp} and studied by several groups worldwide (see \cite{Freidel:2005qe,Oriti:2006se} for reviews and \cite{Magnen:2009at,Freidel:2009hd,Krajewski:2010yq,Ashtekar:2010ve,Oriti:2010hg,Gurau:2010ba,Baratin:2011tg} for a sample of recent results). In three dimensions, it can be described as a scalar field on the homogeneous space $G^3$, where $G$ is a compact Lie group, usually $\textrm{SU}(2)$, whose perturbative expansion generates `stranded' Feynman graphs analogous to the familiar ribbon graphs of matrix models. Just like ribbon graphs are naturally dual to surfaces, these stranded graphs can be associated to topological spaces obtained by gluing tetrahedra along their boundary \cite{DePietri:2000ii}. More precisely, a tetrahedron is associated to each four-valent vertex of the graph, a boundary triangle to each half-edge emanating from it, and a simplicial map between these triangles to each `stranded' edge. The `gluing' defined by such a stranded graph is the quotient of the tetrahedra along the simplicial maps. Similar constructions are widely studied in algorithmic topology \cite{matveev,hatcher,thurston}; it can be showed in particular that all closed three-dimensional manifolds have such a presentation.\footnote{Burton has developed a very useful freeware, Regina, to study three-manifolds in this way \cite{regina}.}

An interesting question raised by Gurau is the possible generalization of the `planar limit' to group field theory, with three-dimensional manifolds of trivial topology dominating the perturbative expansion of the group field theory in a certain regime \cite{Gurau:2010ba}. To address it, a key ingredient is the relationship between the combinatorics of stranded graphs and the topology of the corresponding gluing. This was studied by DePietri and Petronio in \cite{DePietri:2000ii}. In the paper quoted in the title \cite{Gurau:2010nd}, Gurau contradicted their findings and argued that GFT gluings are in general $(i)$ not triangulable and $(ii)$ not pseudo-manifolds. Both claims are erroneous.

Recall that a space is triangulable if it is homeomorphic to the polyhedron of a simplicial complex. It is true that gluings are usually not simplicial complexes, because of tadpoles (two triangles of a single tetrahedron glued together) and multiple edges (two tetrahedra glued along several triangles). Moreover, and more subtly, the simplicial structure of the tetrahedra does not necessarily induce a cellular structure on the quotient: the projection of the simplices to their image in the gluing are not the characteristic maps of a CW complex. This is because the identification of triangles can result in the folding of edges onto themselves, thereby spoiling the injectivity of the projections on the interior of the edges. 

That the gluings are not naturally equipped with a cellular structure, however, does not mean that they are not triangulable. Indeed, they are: it suffices to apply a sufficient number of barycentric subdivisions to the tetrahedra prior to their gluing to make sure that none of the new edges get folded onto themselves, thus inducing a triangulation on the quotient \cite{DePietri:2000ii,thurston}. 

Gurau's second claim is that gluings are not all pseudo-manifolds. Roughly speaking, a pseudo-manifolds is a manifold with singularities of codimension at least two. (Hence, in three dimensions, a pseudo-manifold can be thought of as a manifold `pinched along lines'.) It is considered important for the viability of the group field theory program that it does not generate spaces more singular than that. 

The most useful definition of a pseudo-manifold is combinatorial: a pseudo-manifold is a topological space having a triangulation with the following three properties.\footnote{Notice that from a combinatorial perspective, a pseudo-manifold is a \emph{simpler} object than a manifold.}

\begin{itemize}
\item
It is non-branching: every $(n-1)$-simplex is the face of exactly two $n$-simplices.
\item
It is strongly connected: every pair of $n$-simplices can be connected by a finite sequence of $n$-simplices in which consecutive simplices share an $(n-1)$-simplex.
\item
It is pure: every simplex is the face of some simplex.
\end{itemize}
Of course, since they are not simplicial complexes themselves, one cannot check this definition directly on the gluings. However, it is straightforward to check that their triangulations obtained by subdividing the tetrahedra do satisfy these conditions, and thus that any gluing is a pseudo-manifold. In fact, it was showed in \cite{DePietri:2000ii} that the only potentially singular points in the original tetrahedra are the vertices and the midpoint of edges.

Instead of considering this triangulation (which he claims does not exist), Gurau used the Euler characteristic to check whether gluings are pseudo-manifolds.  Indeed, since the Euler characteristic of a pseudo-manifold is non-negative \cite{thurston}, he argued that if a gluing turns out to have a negative Euler characteristic, then it is not a pseudo-manifold. He then considered several examples, for which he found that the alternating sum $\sum_{j=0}^3(-1)^jc_j$, where $c_j$ is the cardinal of a certain multiset representing the image of $j$-simplices in the gluing, is negative, and thus concluded that they are not pseudo-manifolds. He called this phenomenon ``wrapping singularities", and put forward a modified version of group field theory, coined ``colored" group field theory, to cure this problem.

The catch is that the sum $\sum_{j=0}^3(-1)^jc_j$ is \emph{not} the Euler characteristic of the gluing, because $c_j$ is not the number of $j$-cells in a cell decomposition, for the same reason that the projections of the simplices onto their images in the gluing are not the characteristic map of a CW complex. In fact, the ``wrapping singularities" identified by Gurau express precisely the failure of these maps to be characteristic maps. (For instance, in the example $\mathcal{G}^1$ with a single tetrahedron of \cite{Gurau:2010nd}, the edges $(0,1)$ and $(2,3)$ get folded onto themselves.) The criterion he gives for their avoidance, in terms of the strands of the GFT graph, is nothing but the \emph{compatibility} or \emph{admissibility} condition usually imposed on simplicial gluings to respect their cellular structure \cite{matveev,hatcher,thurston}. 

To conclude, since it is not true that GFT gluings are not pseudo-manifolds, we cannot agree with Gurau that ``the situation looks bleak for [standard, that is non-colored] GFT" \cite{Gurau:2010nd}. Or if it does, it must be for other reasons. 

\begin{acknowledgements}

I thank Razvan Gurau for several discussions on group field theory and topology.

\end{acknowledgements}
%%%%%%%%%%%%%%%%%%%%%%%%%%%%%%%%%%%%%%%%%%%%%%%%%%%
%%%%%%%%%%%%%%%%%%%%%%%%%%%%%%%%%%%%%%%%%%%%%%%%%%%
\bibliographystyle{utcaps}

\bibliography{Bibliographie}

%%%%%%%%%%%%%%%%%%%%%%%%%%%%%%
%%%%%%%%%%%%%%%%%%%%%%%%%%%%%%

\end{document}